\pdfoutput=1

\RequirePackage{fix-cm}

\documentclass[twocolumn,numbers,sort&compress]{svjour3}

\smartqed

\usepackage{graphicx}
\usepackage{amssymb}
\usepackage{amsmath}
\usepackage{txfonts}
\usepackage{natbib}
\usepackage{hyperref}
\usepackage{color}

\hypersetup{pdftitle = The title of my PDF, pdfauthor = My name, pdfsubject= The subject, pdfkeywords = keyword1 keyword2 keyword3}
\hypersetup{colorlinks = true, linkcolor = blue, anchorcolor = red, citecolor = blue, filecolor = red, pagecolor = red, urlcolor = blue}

\journalname{}

\begin{document}

\title{On the dynamics of a seventh-order generalized H\'{e}non-Heiles potential}

\author{Fredy L. Dubeibe  \and Euaggelos E. Zotos \and Wei Chen}

\institute{
  F. L. Dubeibe \at
    Facultad de Ciencias Humanas y de la Educaci\'on, \\
    Universidad de los Llanos, Villavicencio, Colombia.\\
    Corresponding's author email: \url{fldubeibem@unal.edu.co}
  \and
  E. E. Zotos \at
    Department of Physics, School of Science, \\
    Aristotle University of Thessaloniki, \\
    GR-541 24, Thessaloniki, Greece \\
    \email{\url{evzotos@physics.auth.gr}}
  \and
  W. Chen \at
    LMIB \& School of Mathematics and Systems Science, \\
    Beihang University, Beijing 100191, China\\
    Peng Cheng Laboratory, Shenzhen, \\
    Guangdong, China, 518055
    Beijing Advanced Innovation Center for Big Data and Brain Computing, \\
    Beihang University, Beijing 100191, China \\
    \email{\url{chwei@buaa.edu.cn}}
}

\date{Received: - / Accepted: - / Published online: -}

\titlerunning{On the dynamics of a seventh-order generalization of the H\'{e}non-Heiles potential}

\authorrunning{Dubeibe et al.}

\maketitle

\begin{abstract}

This paper deals with the derivation and analysis of a seventh-order generalization of the H\'enon-Heiles potential. The new potential has axial and reflection symmetries, and finite escape energy with three channels of escape. Based on SALI indicator and exits basins, the dynamic behavior of the seventh-order system is investigated qualitatively in cases of bounded and unbounded movement. Moreover, a quantitative analysis is carried out through the percentage of chaotic orbits and the basin entropy, respectively. After classifying large sets of initial conditions of orbits for several values of the energy constant in both regimes, we observe that when the energy moves away from the critical value, the chaoticity of the system decreases and the basin structure becomes simpler with sharper and well defined bounds. Our results suggest that when the seventh-order contributions of the potential are taken into account, the system becomes less ergodic in comparison with the classical version of the H\'enon-Heiles system.


\end{abstract}

\section{Introduction}
\label{intro}

In 1963 Michael H\'enon and Carl Heiles, numerically investigated the existence of the third integral of motion in axisymmetric potentials \cite{HH64}, motivated by the apparent existence of such quantity in a large number of galactic orbits. This system is usually called the H\'enon-Heiles Hamiltonian and, to our knowledge, it was the first one for which a third integral of motion was constructed \cite{C60}. Although its first application was the explanation of the non-equality on the axes of the velocity ellipsoid for stars in the solar neighborhood, its current uses range from the calculation of energies and lifetimes for metastable states within the quantum mechanics formalism \cite{WM81}, to the analysis of gravitational waves, by means of general relativity \cite{K98}.

Exploiting the properties of Hamiltonian systems with two degrees of freedom (see e.g. \cite{BBS10}), over the last few decades many real problems coming from Physics and Chemistry have been formulated using modifications to the H\'enon-Heiles potential. For example, the chaotic ionization mechanism in chemical reactions is modeled with a rotating potential of H\'enon–Heiles type \cite{RPBF09}, laser-driven reactions with several open channels are modeled using a driven H\'enon-Heiles system \cite{KBJBPU07}, the description of motion of test particles in presence of vacuum gravitational pp-wave spacetimes was studied using a modified H\'enon-Heiles potential \cite{VP00}, or the modeling of black holes with external halos is realized through the superposition of quadrupolar and octupolar terms, such that in the Newtonian limit, the gravitational interaction reduces to an analog of the H\'enon-Heiles potential \cite{VL96}, just to name a few.

On the other hand, many efforts have been made to generalize the H\'enon-Heiles potential, keeping its main characteristic intact, i.e., being a time-independent two-dimensional Hamiltonian system with finite escape energy. The very first generalization was performed by Verhulst \cite{V79}, who extended up to the fourth-order this potential, to produce qualitative and quantitative results for the study of the resonance cases 1:2, 1:1, 2:1 and 1:3. More recently, the H\'enon-Heiles potential was extended up to the fifth-order, where the authors studied the basins of convergence of equilibrium points of the new system \cite{ZRD18} and carried out a dynamical analysis of unbounded and bounded movement \cite{DRZ18}. 

Our goal in this paper is twofold. First, we shall perform a Taylor series expansion up to the seventh-order of a general form of a potential with reflection and axial symmetries, such that the derived Hamiltonian reduces to the lower order cases: H\'enon-Heiles, Verhulst and the fifth-order system. Second, by using some state-of-the-art numerical techniques like SALI indicators and exits basins, the dynamic behavior of the new system is investigated in the cases of open and closed zero-velocity curves. The quantitative analysis is carried out through the percentage of chaotic orbits and the basin entropy, respectively.

The present paper is organized as follows. In section \ref{derpot}, the derivation of the generalized potential along with the Hamiltonian and equations of motion are presented. In section \ref{result}, the existence and stability of the fixed points are investigated by means of the standard linear stability analysis. Next, we describe the methodology and techniques used to obtain the numerical results for bounded and unbounded orbits. Also, we show how the configuration $(x,y)$ and the $(x,E)$ planes evolve, as a function of a perturbation parameter. Finally, in section \ref{conc} we present the main conclusions and implications for the complex systems field, derived from this work.

\section{Derivation of the Potential}
\label{derpot}

Let us consider the case of a test particle in the presence of an axially symmetric potential. Given the symmetry properties of the system, the usual and appropriate coordinate system of choice is cylindrical $(r, \theta, z)$. Moreover, the symmetry with respect to the rotational degree of freedom gives place to a cyclic coordinate and therefore, the Hamiltonian can be written as
\begin{equation}
{\cal H} = \frac{1}{2}(\dot{r}^{2} + \dot{z}^{2}) + \Phi_{\rm{eff}}(r,z),
\label{ham0}
\end{equation}
with $\Phi_{\rm{eff}}(r,z) = \Phi (r, z) + L_{z}^{2}/2 r^2$, the effective potential and $L_z$, the angular momentum along the $z-$axis.

Following Contopolulos \cite{C02}, it is possible to find an approximate analytical form of the axisymmetric potential by performing a Taylor series expansion of $\Phi_{\rm{eff}}(r,z)$ around $(r_{0},0)$ up to the 7th-order. Under the following considerations: {\it (i)} The effective potential $\Phi_{\rm{eff}}(r,z)$ has a minimum at $(r,z)=(r_{0},0)$, where $\partial \Phi/\partial r=L_{z}^{2}/r^{3}$, and {\it (ii)} By imposing reflection symmetry about the $z = 0$ plane, the potential must contain only even powers on $z$ and hence the odd-powered derivatives of $\Phi (r, z)$ are odd functions, the approximate potential can be simplified to
\begin{eqnarray}
\Phi_{\rm eff}(r,z) &&\approx  a_{1} \xi ^4+z^4 \left(a_{2}+b_{2} \xi +c_{3} \xi^2+d_{4} \xi ^3\right)+z^2\nonumber\\
   &&\times \left(a_{3} \xi ^2+b_{3} \xi ^3+c_{4} \xi ^4+d_{3} \xi ^5+\omega_{2}^2+\xi\epsilon \right)+\beta  \xi ^3\nonumber\\
   &&+b_{1} \xi ^5+c_{1} \xi^6+z^6 (c_{2}+d_{2} \xi)+d_{1} \xi ^7+\xi ^2
 \omega_{1}^2,
\label{expan}
\end{eqnarray}
where $\xi = r - r_0$, we omitted constant terms, and the terms $a_{i}, b_{i}, c_{i}, d_{i}, \omega_{i}, \epsilon$, and $\beta$ are constants.

Replacing $z\rightarrow x$, $\xi\rightarrow y$, and setting $a_{1}$ = $a_{2}$ = $\delta/4$, $a_{3}$ = $c_{3}$ = $c_{4}$ = $\delta/2$, $b_{2}$ = $b_{3}$ = $d_{2}$ = $d_{3}$ = $d_{4}$ = $\delta$, $c_{1}$ = $c_{2}$ = $\delta/6$, $\epsilon$ = $1$, $b_{1}$ = $-\delta/5$, $d_{1}$ = $-\delta/7$, $\beta$ = $-1/3$, and $\omega_{1}$ = $\omega_{2}$ = $1/\sqrt{2}$, the final expression for the effective potential read as
\begin{eqnarray}
\Phi_{\rm{eff}}(x,y)& = &\frac{1}{2} \big(x^2 + y^2\big) + x^2 y - \frac{y^3}{3} + \delta \Bigg[x^6 y + x^4 y^3 + x^4 y\nonumber \\
&&  + x^2 y^5 + x^2 y^3 - \frac{y^7}{7} - \frac{y^5}{5}\nonumber\\
&& + \frac{1}{4} \Big(x^2 + y^2\Big)^2 + \frac{1}{6} \Big(x^2 + y^2\Big)^3 \Bigg]\,.
\label{pot}
\end{eqnarray}

Let us enumerate the main properties of the potential presented above:
\begin{enumerate}
\item It has finite escape energy, such that after some energy threshold the zero-velocity curves (henceforth ZVC) exhibit three channels of escape.
\item By setting $\delta = 0$ in Eq. (\ref{pot}), it reduces to the well known classical H\'enon-Heiles potential.
\item With an appropriate setting of the constant parameters, it reduces to the fourth-order \cite{V79} and fifth-order \cite{DRZ18} generalizations of the H\'enon-Helies potential.
\end{enumerate}

\begin{table}
\caption{Critical energy values for bounded motion $E < E_{\rm{min}}$ and unbounded motion, with three exit channels $E > E_{\rm{max}}$.}
\label{tab1}
\begin{tabular}{lll}
\hline\noalign{\smallskip}
$\delta$ & $E_{\rm{min}}$ & $E_{\rm{max}}$  \\
\noalign{\smallskip}\hline\noalign{\smallskip}
0.0 & 0.166666666666666685 & 0.166666666666666685 \\
0.1 & 0.150995534705215761 & 0.174047619047619062 \\
0.2 & 0.143035865330651490 & 0.181428571428571439 \\
0.3 & 0.137831022344923199 & 0.188809523809523816 \\
0.4 & 0.134069419068656492 & 0.196190476190476193 \\
0.5 & 0.131196376439182749 & 0.203571428571428570 \\
0.6 & 0.128923859750555619 & 0.210952380952380947 \\
0.7 & 0.127082992635008080 & 0.218333333333333324 \\
0.8 & 0.125566355505970118 & 0.225714285714285701 \\
0.9 & 0.124301474019574282 & 0.233095238095238078 \\
1.0 & 0.123237212441469735 & 0.240476190476190455 \\
\noalign{\smallskip}\hline
\end{tabular}
\end{table}

\begin{figure*}[!t]
\centering
\resizebox{\hsize}{!}{\includegraphics{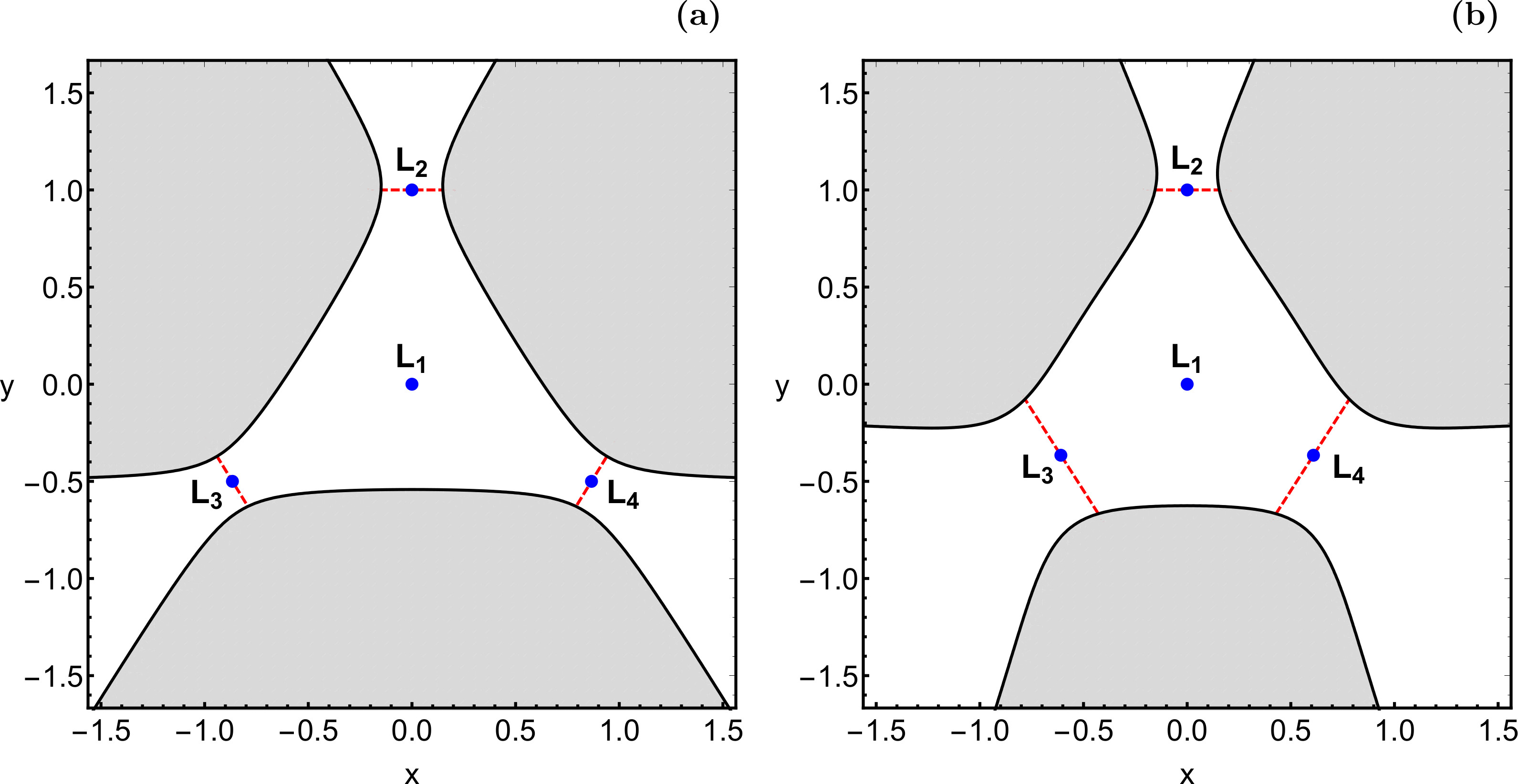}}
\caption{The ZVC on the configuration $(x,y)$ plane, when (a): $\delta = 0$, $E = 0.2$ and (b): $\delta = 1$, $E = 0.35$. The positions of the points of equilibrium are denoted using blue dots, while with red dashed lines we depict the Lyapunov orbits of the system. (Color figure online).}
\label{conts}
\end{figure*}

\begin{figure}[!t]
\centering
\resizebox{\hsize}{!}{\includegraphics{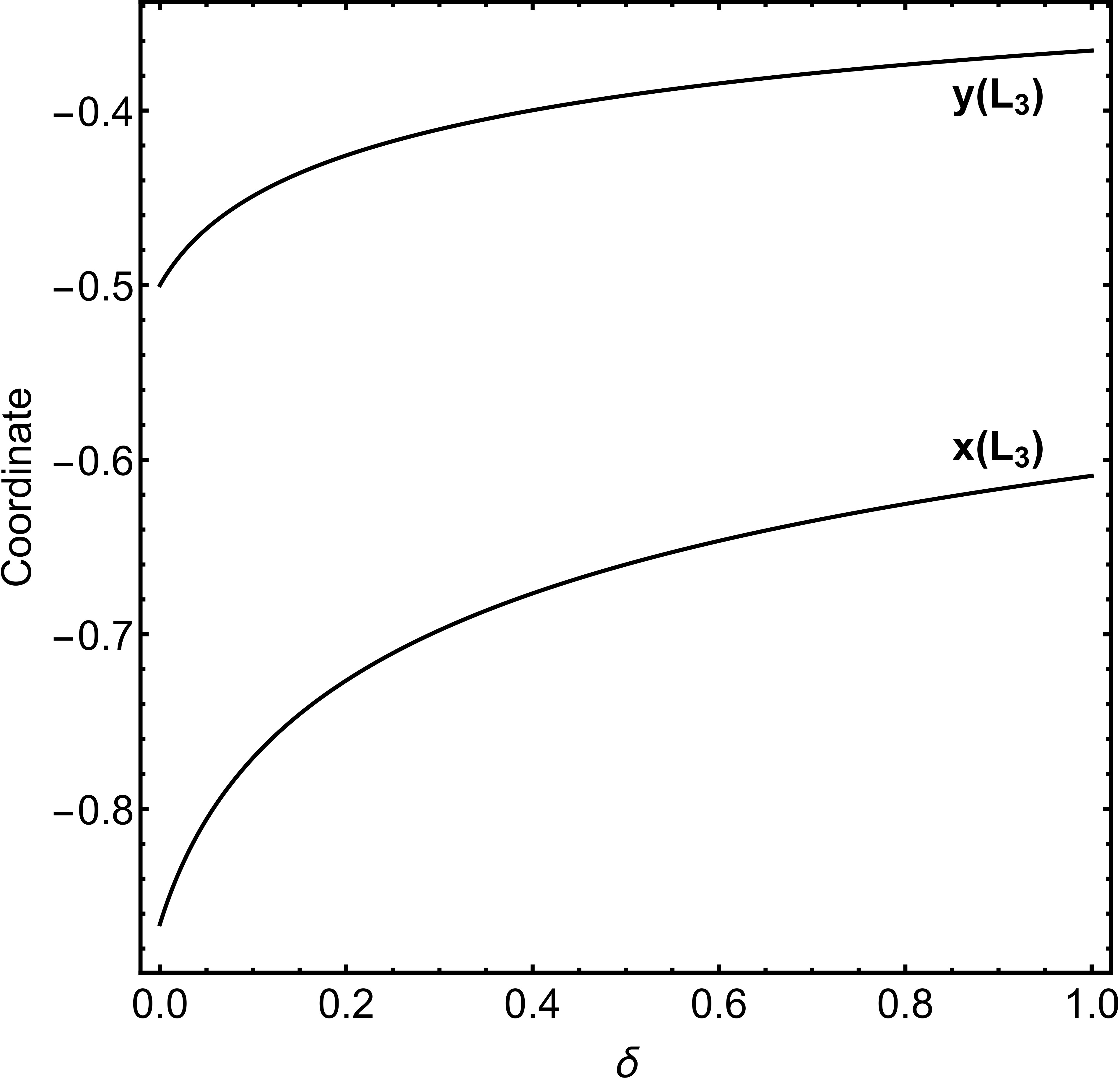}}
\caption{Evolution of the $x$ and $y$ coordinates of the equilibrium point $L_3$, as a function of the parameter $\delta$.}
\label{evol}
\end{figure}

On the other hand, given the dependence of the Hamiltonian with the velocities, the equations of motion can be written in compact form as
\begin{eqnarray}
\ddot{x}& = &- \frac{\partial \Phi_{\rm{eff}}}{\partial x},\\
\ddot{y}& = &- \frac{\partial \Phi_{\rm{eff}}}{\partial y}.
\end{eqnarray}

Taking into account that the system is scleronomous and there are no velocity-dependent terms in the potential energy, we can write ${\cal H} = E$, in such a way that the orbits must be restricted to the region satisfying
\begin{eqnarray}
E&\ge& \frac{1}{2} \big(x^2 + y^2\big) + x^2 y - \frac{y^3}{3} + \delta \Bigg[x^6 y + x^4 y^3 + x^4 y\nonumber \\
&&+ x^2 y^5 + x^2 y^3 - \frac{y^7}{7} - \frac{y^5}{5} + \frac{1}{4} \Big(x^2 + y^2\Big)^2\nonumber\\
&&+ \frac{1}{6} \Big(x^2 + y^2\Big)^3 \Bigg].
\label{reg}
\end{eqnarray}

The escape energy $E_{\rm{esc}}$, is defined as the smallest energy value for which the ZVC open and the test particle is allowed to escape to infinity. In the present study we are interested in bounded and unbounded motion, however, aiming to compare with the H\'enon-Heiles system in which energies above $E > E_{\rm{esc}}$ allows the test particle to escape to infinity through three different channels, in Table \ref{tab1} we present the critical values of energy for different values of $\delta$ satisfying both requirements, i.e., $E_{\rm{min}}$ refers to the largest energy value for which the ZVC are closed, while $E_{\rm{max}}$ defines the smallest energy value for which the ZVC has three escape channels.

\section{Results}
\label{result}

We start by considering the fixed points of the system and their stability. The system always exhibits four equilibrium points, regardless of the value of $\delta$, such equilibriums shall be identified as $L_1$, $L_2$, $L_3$, and $L_4$. In Fig.~\ref{conts} we present the open ZVC of the system, along with the positions of the equilibrium points for $\delta = 0$ and $\delta = 1$, respectively. At the same diagram, we also depict the positions of the Lyapunov orbits \cite{L07,L49}, which control the escape of the particles. When $\delta$ varies in the interval $[0,1]$ the fixed point $L_1$ keeps at the origin $(0,0)$, while $L_2$ stays static at $(0,1)$. Only the equilibrium points $L_3$ and $L_4$ change position. In Fig.~\ref{evol} we show the parametric evolution of the coordinates $x(L_3)$ and $y(L_3)$, as a function of $\delta$. The evolution suggests that for increasing values of $\delta$ the equilibrium points $L_3$ and $L_4$ move toward the origin, following a nonlinear path. The positions of $x(L_4)$ and $y(L_4)$ are not presented because they are symmetric with respect to the $y-$axis, i.e., $x(L_4)= - x(L_3)$ and $y(L_4) = y(L_3)$. Concerning the stability, it is found that all libration points are always linearly unstable, when $\delta \in [0,1]$.

In what follows, we analyze the dynamics of the seventh order generalized H\'enon-Heiles potential. In general terms, we shall classify the bounded and unbounded orbits by using color coded diagrams to distinguish between the different types of the orbits, i.e. for closed ZVC, the classification corresponds to regular, chaotic or sticky orbits, while for open ZVC, the dynamic behavior is richer adding three additional final states, depending on the exit channels: exit via channel 1, exit via channel 2 and exit via channel 3. Once the classification is made, we will use quantitative indicators to inspect the evolution of chaoticity in comparison with the classical H\'enon-Heiles system. Let us start with the case in which energy demands all orbits to be bounded.

\subsection{Bounded orbits}

The dynamics of the system in the case of closed ZVC is studied in the present subsection. Here, aiming to reveal the differences in the dynamical behavior of the seventh-order generalization and the H\'enon-Heiles system, we shall use the chaos indicator SALI \cite{S01}. In particular, we use different values of $\delta$, allowing us to gradually control the contribution of higher-order terms in the system. Moreover, we decrease the energy with respect to its respective critical value to see the influence on the global dynamics of the system.

\begin{figure*}[!t]
\centering
\resizebox{\hsize}{!}{\includegraphics{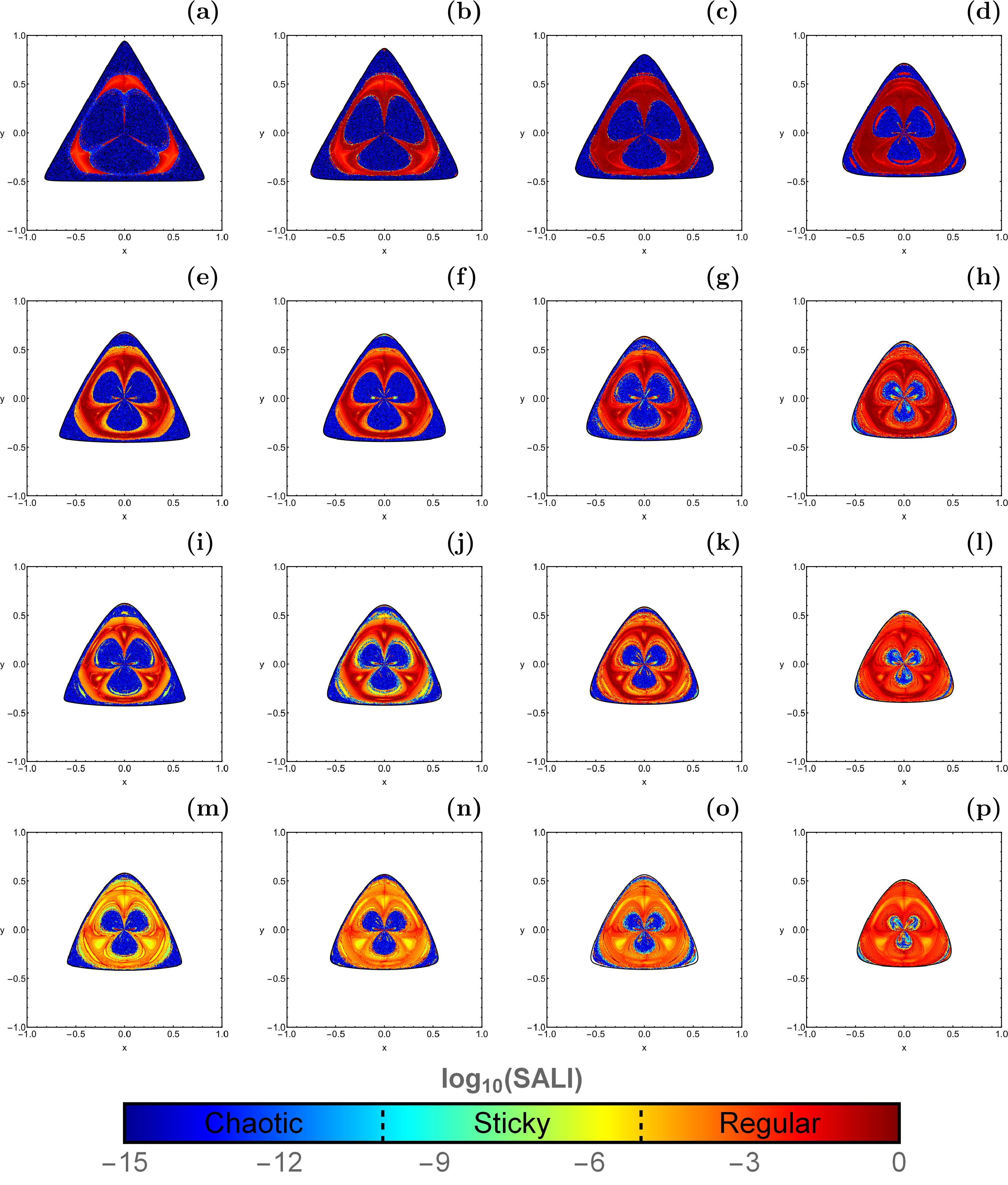}}
\caption{The value of SALI for different values of $\delta$ and energy $E$. The energy decreases according to the formula $E = E_{\rm min}(1 - n/100)$ with $E_{\rm min}$ given in Table \ref{tab1}. Each panel corresponds to the following pairs ($\delta, n$): (a): (0, 1), (b): (0, 5), (c): (0, 10), (d): (0, 20), (e): (0.3, 1), (f): (0.3, 5), (g): (0.3, 10), (h): (0.3, 20), (i): (0.6, 1), (j): (0.6, 5), (k): (0.6, 10), (l): (0.6, 20), (m): (1, 1), (n): (1, 5), (o): (1, 10), (p): (1, 20). (Color figure online).}
\label{sali}
\end{figure*}

\begin{figure}[!t]
\centering
\resizebox{\hsize}{!}{\includegraphics{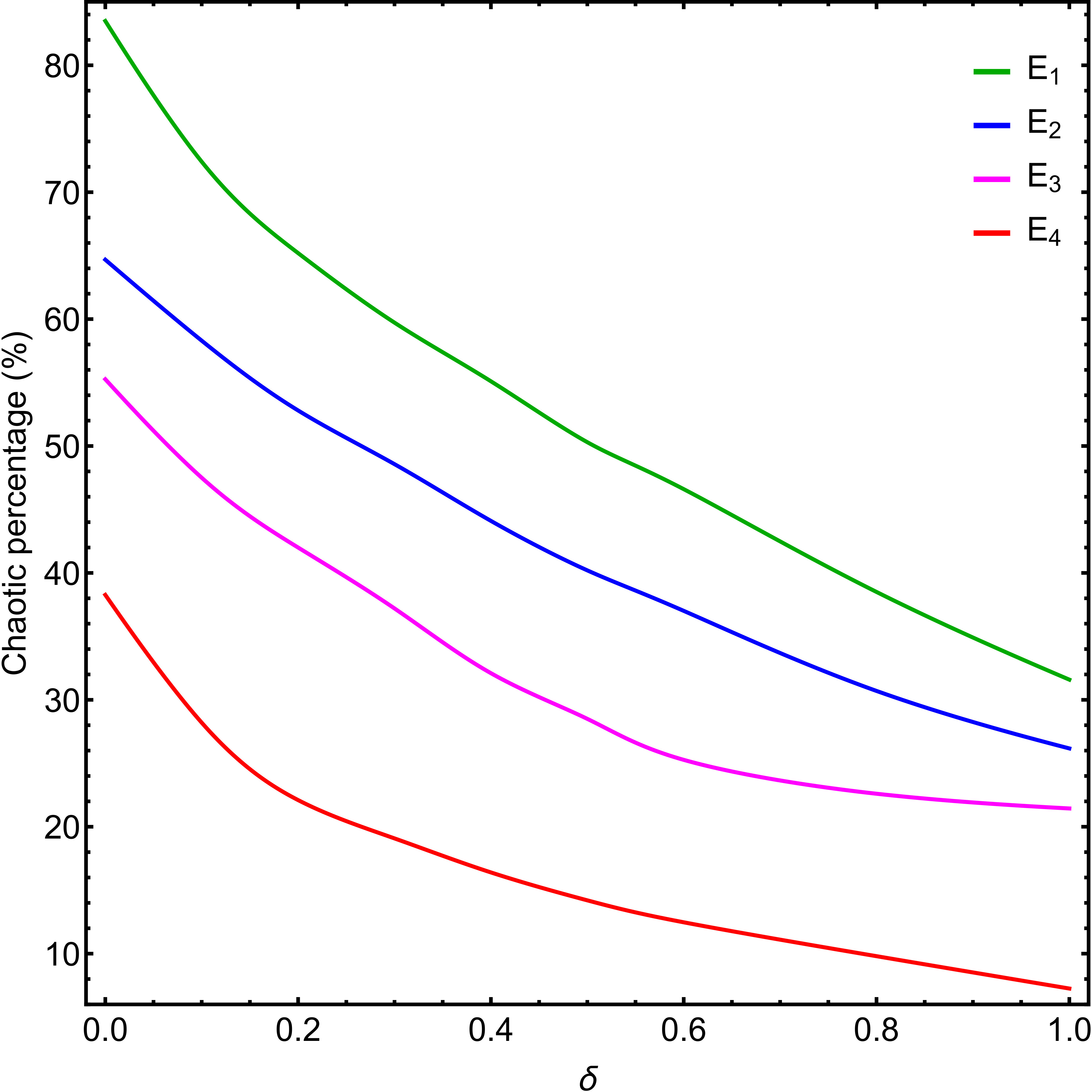}}
\caption{Evolution of the percentage of chaotic orbits, as a function of the parameter $\delta$. (Color figure online).}
\label{cha}
\end{figure}

\begin{figure*}[!t]
\centering
\resizebox{\hsize}{!}{\includegraphics{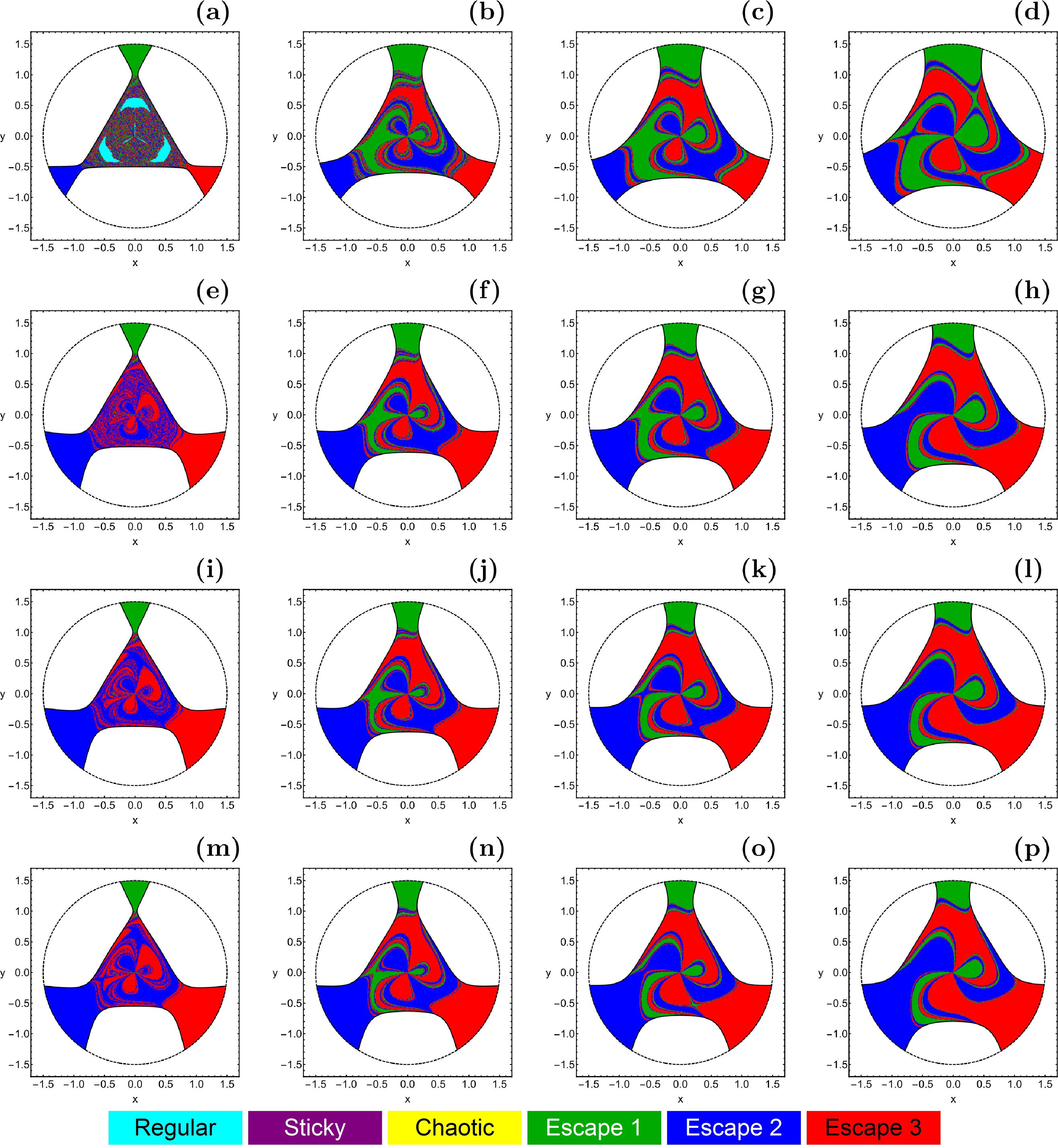}}
\caption{Exit basins for different values of $\delta$ and energy. The energy increases according to the formula $E = E_{\rm max}(1 + n/100)$ with $E_{\rm max}$ given in Table \ref{tab1}. Each panel corresponds to the following pairs ($\delta, n$): (a): (0, 2), (b): (0, 50), (c): (0, 100), (d): (0, 200), (e): (0.3, 2), (f): (0.3, 50), (g): (0.3, 100), (h): (0.3, 200), (i): (0.6, 2), (j): (0.6, 50), (k): (0.6, 100), (l): (0.6, 200), (m): (1, 2), (n): (1, 50), (o): (1, 100), (p): (1, 200). (Color figure online).}
\label{esc}
\end{figure*}

\begin{figure*}[!t]
\centering
\resizebox{0.8\hsize}{!}{\includegraphics{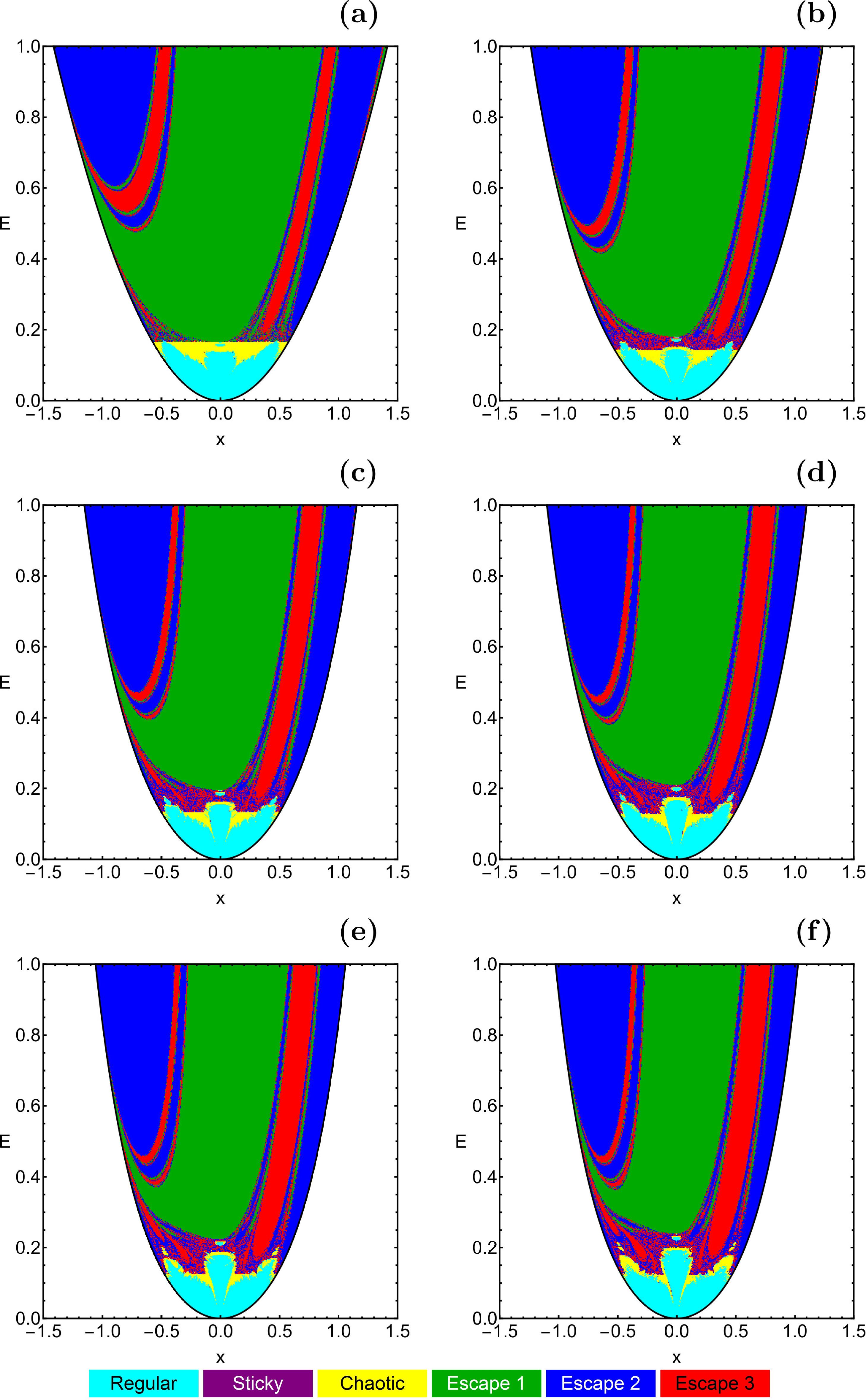}}
\caption{Orbit classification on the $(x,E)$-plane for increasing values of the parameter $\delta$. (a): $\delta = 0$, (b): $\delta = 0.2$, (c): $\delta = 0.4$, (d): $\delta = 0.6$, (e): $\delta = 0.8$, and (f): $\delta = 1$. (Color figure online).}
\label{xE}
\end{figure*}

\begin{figure}[!t]
\centering
\resizebox{\hsize}{!}{\includegraphics{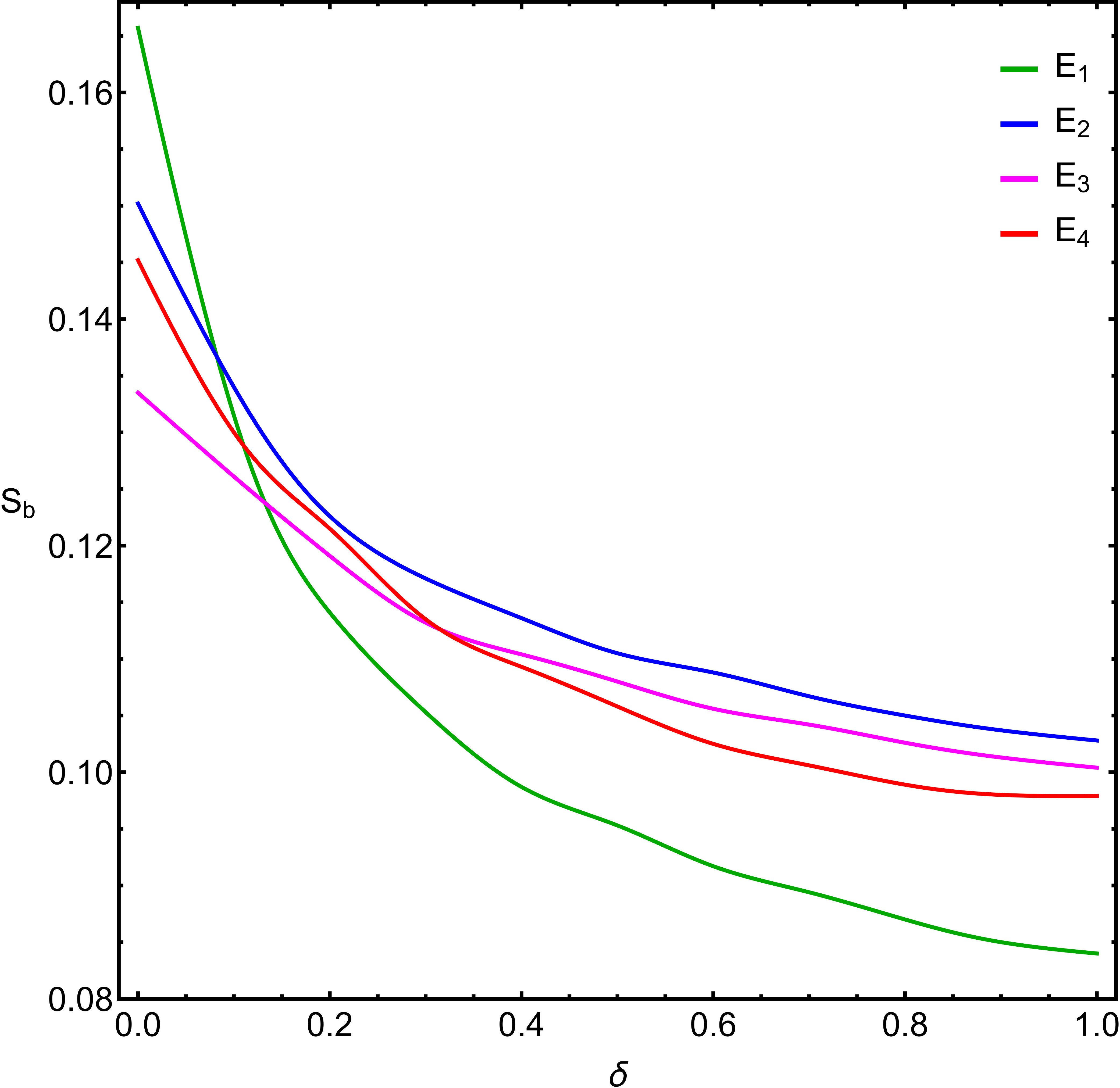}}
\caption{Basin entropy for the exit basins, as a function of the parameter $\delta$, using fixed values of energy. (Color figure online).}
\label{Sb}
\end{figure}

For a full overview of the dynamics, we define a dense grid of $1024 \times 1024$ initial conditions, uniformly distributed and closely spaced on the configuration plane $(x,y)$. The initial conditions are set in such a way that they satisfy the energy restrictions, i.e., for each initial coordinate position $(x_0, y_0)$ with $\dot{y}_{0} = 0$ the initial condition for $\dot{x}_{0} = 0$ is always defined by the relation ${\cal H} = E$. All our numerical calculations were performed with a double-precision adaptive Bulirsch-Stoer algorithm, implemented in \verb!FORTRAN 77! \cite{PTFV92}, where it was found that by decreasing the magnitude of the time step to $10^{-2}$, the numerical error has been reduced to an approximate order of $10^{-13}$, or smaller. It is important to note that we use SALI for distinguishing between chaotic and ordered motion instead of any other indicator (e.g. GALI, LCN, FLI, etc.) because it is not only fast and easy to compute but also efficient and very accurate. According to the literature, orbits are classified as regular if the value of SALI for a maximum integration time ($10^4$ in our case) is larger than $10^{-4}$, or chaotic if the resulting value is smaller than $10^{-8}$, however, if the final value for SALI belongs to the interval $10^{-8} \le {\rm{SALI}} \le 10^{-4}$ the orbit is classified as sticky. In this context, sticky is referred to an orbit that needs a larger integration time in order to reveal its chaotic nature\footnote{We refer the interested reader to \cite{S01}.}.

In Fig.~\ref{sali} we classify each orbit according to SALI, for different values of $\delta$ and energy $E$. Each row in Fig.~\ref{sali} has a different value of $\delta$, from up to down: 0, 0.3, 0.6, 1. Also, each column corresponds to a different value of energy varying according to the expression $E = E_{\rm min}(1 - n/100)$, such that we reduce the total energy in $n\%$ with respect to the critical energy, in particular, $n$ takes (from left to right) the values: 1, 5, 10, 20. The color code used for the final value of SALI is presented in the lower part of Fig.~\ref{sali}, where the rainbow color palette ranges from dark blue (SALI = $10^{-15}$) to dark red (SALI = $\sqrt{2}$).

When reading the figure from left to right it is found that as the energy becomes smaller than the critical energy $E_{\rm{min}}$, the area occupied by the regular regions increases in comparison with the area of the chaotic orbits. This behavior has been studied extensively in the theoretical and numerical regimes for the H\'enon-Heiles system (see e.g. \cite{KR92}), but there seems to be also present for $\delta\in (0,1]$. Furthermore, when reading the figure from up to down, it is observed that increasing the values of the $\delta$ parameter reinforces the tendency of the system to increase the area occupied by regular orbits, or in other words, from a qualitative point of view, the amount of chaotic orbits seems to decrease as $\delta$ increases.

A quantitative measure of the percentage of chaotic orbits is presented in Fig.~\ref{cha}. The proportion of chaotic orbits (relative area covered by the blue pixels) is calculated as a function of the parameter $\delta$, using four energy levels, $E_1$, $E_2$, $E_3$, and $E_4$, which correspond to $n = 1$, $n = 5$, $n = 10$, and $n = 20$, respectively. In the four cases we observe that when $\delta$ tends to 1 (additional terms equally weighted), the percentage of chaotic orbits is reduced in approximately 40\% of its initial value for $\delta = 0$. Additionally, it can be easily noted that the percentage of regular orbits approximately reach the hundred percent for $\delta = 1$ and $E = E_4$. These results suggest that when the seventh order contributions of the potential are taken into account, the system becomes less ergodic in comparison with the classical version of the H\'enon-Heiles system and in some cases practically regular.

\subsection{Unbounded orbits}

The dynamic analysis of the system in the case of open ZVC with escape channels is studied in the present subsection. In this case, the exit basins are plotted using energy values beyond the threshold limit $E_{\rm max}$ which directly depends on the $\delta$ parameter, as shown in Table \ref{tab1}. In general, exit basins are defined as the set of initial conditions that escape through a particular exit, hence, the complexity of the basins may depend on the number of exits on the ZVC. However, as in the present study the system always exhibits three channels of escape, the structure of the exit basins shall be related uniquely to the effect of the $\delta$ parameter and the total energy of the system $E$.

The initial conditions for each orbit are set by defining a position coordinate ${\bf{r}} = (x,y)$, while the initial velocities are defined by ${\bf{r}}\cdot{\bf{\dot{r}}} = 0$ and $E = {\cal H}$, with ${\bf{r}}\times{\bf{\dot{r}}} > 0$. For the classification of the orbits we use the standard convention, i.e., if the test particle escapes through exit $i$, its initial conditions will belong to the exit $i$-th basin. The exits present in our system are denoted as follows: a particle escaping to infinity through $y\rightarrow \infty$ belongs to the channel 1 (green), if $(x\rightarrow - \infty, y\rightarrow - \infty)$ belongs to the channel 2 (red), or if $(x\rightarrow \infty, y\rightarrow - \infty)$ belongs to the channel 3 (red). It is a well-known fact that for energy values larger than the escape energy a portion of orbits does not escape from the system, and sometimes forming islands around the stable periodic orbits \cite{B84}, this phenomenon occurs mainly for energy levels, just above the critical value of the energy of escape.

In Fig.~\ref{esc} we present the exit basins obtained for four different values of $\delta$ and energy. Each row has a different value of $\delta$, from up to down: 0, 0.3, 0.6, 1, while, each column corresponds to a different value of energy varying according to the expression $E = E_{\rm max}(1 + n/100)$, such that $n$ corresponds to the percantage increase in energy with reference to the critical value $E_{\rm max}$ and takes the values (from left to right): 1, 5, 10, and 20. The first row corresponds to the H\'enon-Heiles system, where it can be noted that for larger percentages of deviation in respect of $E_{\rm max}$, the basin boundaries become smoother and also well-defined, in fact, for $n \ge 50$ the amount of orbits escaping through the channels is so high that the bounded orbits become imperceptible to the naked eye. The same behavior is perceived for $\delta \in (0,1]$ (second, third and fourth row), nevertheless, unlike the previous case, the lower exit channels show a symmetric wider bottleneck thus allowing most orbits to go to infinity through these two channels.

The global dynamical behavior of the system can be analyzed by means of the orbit classification in the $(x, E)$ plane. In Fig.~\ref{xE} we present the final states of the system using a continuous spectrum of values of energy, for six values of $\delta$ = 0, 0.2, 0.4, 0.6, 0.8, and 1. The initial conditions on $x$-position and energy are selected from the allowed regions of motion, while $y = \dot{x} = 0$ and $\dot{y}$ is determined by the energy conservation $E = {\cal H}$. This figure shows that for larger values of $E$ the phase space is mainly occupied by escaping orbits. For energy values close to $E_{\rm min}$, practically all the final states are possible but the basin boundaries are poorly defined and the extent of these regions increases with the increase in the parameter $\delta$. Finally, when $E\rightarrow 0$ the orbits of the system are predominantly regular with a chaotic layer surrounding it, however, this region becomes less uniform as $\delta \rightarrow 1$.

On the other hand, a quantitative measure of the uncertainty of the exit basins is provided by the basin entropy \cite{DWGGS16}. Such an indicator gives us information about the topology of the basins rather than of the evolution of the orbits, or in other words, a measure of the difficulty to predict in advance which will be the final state of an orbit. In a generic case, for a dynamical system with $N_A$ final states, whose phase space was subdivided into a numerical grid of $N$ square boxes, the basin entropy can be calculated as the average of the Gibbs entropy of every box, i.e.,

\begin{equation}
S_{b} = \frac{1}{N}\sum_{i=1}^{N}\sum_{j=1}^{N_A}P_{i,j}\log\left(\frac{1}{P_{i,j}}\right),
\end{equation}
where $P_{i,j}$ represents the probability to get the $j$-th final state inside the $i$-th box\footnote{For details of the theory of the basin entropy, we refer the reader to \cite{DWGGS16}.}.

In Fig.~\ref{Sb} we plot the basin entropy as a function of the $\delta$ parameter using four levels of energy, $E_1$, $E_2$, $E_3$, and $E_4$, which correspond to $n = 2$, $n = 50$, $n = 100$, and $n = 200$, respectively. In all cases, we observe that as the parameter $\delta$ increases, the basin entropy $S_b$ reduces, such decrease is more marked for small $\delta$ values and becomes less pronounced (with a slope close to zero) for high $\delta$ values. Also, it is important to note, that there is no strict relationship for fixed values of $\delta$ and decreasing energies, for example it is observed that when $\delta = 0$ (H\'enon-Heiles system) the basin entropy is larger for levels $E_4$ than for levels $E_3$, or even for a full contribution of the seventh order terms $\delta = 1$ the basin entropy is much smaller for levels $E_1$ than for levels $E_2$.

\section{Conclusions}
\label{conc}

In this paper, by using some state-of-the-art numerical techniques, we have performed a numerical investigation of the dynamics of bounded and unbounded orbits with three channels of escape for a seventh-order generalization of the H\'enon-Heiles potential. Our dynamical model can be used, for example, to describe the motion of stars in the vicinity of the galactic center, since it comes from the Taylor series expansion of an axially symmetric galactic potential, with reflection symmetry about the plane $z=0$.

A systematic classification of orbits is carried out, so that the different types of orbits can be arranged into groups, according to the structure of the zero-velocity curves. For closed contours the orbits are classified as regular, chaotic or sticky, while for open contours with three channels of escape, orbits are assorted as regular, chaotic or sticky, escaping through exit 1, escaping through exit 2 or escaping through exit 3. The method used to determine the chaoticity or regularity of the orbits is the standard SALI method, which provides a reliable and fast indicator of chaos in Hamiltonian systems.

To obtain a graphical representation of the dynamics, we define a dense grid of initial conditions, uniformly distributed and closely spaced on the configuration plane $(x,y)$. The qualitative information supplied by the basins suggests that, in accordance with the well-known results for the classical H\'enon-Heiles Hamiltonian, for energy values larger or smaller than the critical value, the chaoticity of the system decreases and the structure becomes simpler with sharper and well-defined bounds. This fact is confirmed when plotting the basins on the $(x,E)$ plane, thus using a continuous spectrum of energy levels.

Concerning the effect of the seventh-order terms in the orbital dynamics of the system, we studied the quantitative evolution of the percentage of chaotic orbits and the basin entropy for the exit basins, as a function of a parameter $\delta$. Our results are conclusive, pointing out that the percentages of chaotic orbits are significantly reduced for the new potential in comparison with the results for the classical H\'enon-Heiles, even, for very small energy values, the new system approximately reaches the hundred percent or regular orbits. Similarly, the highly fractal basin boundaries of the classical H\'enon-Heiles become smoother and well-defined when a full contribution of the seventh order terms is present. We hope that our results will provide a reference for future research on the field of complex systems, mainly in those studies in which a third integral of motion exists in axisymmetric potentials.

\begin{acknowledgements}
This work was partially supported by  COLCIENCIAS (Colombia) Grant 8863 and by Universidad de los Llanos.
\end{acknowledgements}

\section*{Compliance with Ethical Standards}
\footnotesize

The authors declare that they have no conflict of interest.


\begin{thebibliography}{}

\bibitem{HH64} H\'{e}non, M., Heiles, C.: The applicability of the third integral of motion: some numerical experiments. Astron. J. \textbf{69} 73-79 (1964).

\bibitem{C60} Contopoulos, G.: A third integral of motion in a galaxy. Zeitschrift fur Astrophysik. {\bf 49} 273 (1960).

\bibitem{WM81} Waite, B. A., Miller, W. H.: Mode specificity in unimolecular reaction dynamics: The H\'{e}non-Heiles potential energy surface. The Journal of Chemical Physics. {\bf 74(7)} 3910-3915 (1981).

\bibitem{K98} Kokubun, F.: Gravitational waves from the Henon-Heiles system. Physical Review D. {\bf{57(4)}} 2610 (1998).

\bibitem{BBS10} Barrio, R., Blesa, F., Serrano, S.: Bifurcations and chaos in Hamiltonian systems. International Journal of Bifurcation and Chaos. {\bf 20(05)} 1293-1319 (2010).

\bibitem{RPBF09} Ramilowski, J. A., Prado, S. D., Borondo, F., Farrelly, D.: Fractal Weyl law behavior in an open Hamiltonian system. Phys. Rev. E. {\bf 80} 055201(R) (2009).

\bibitem{KBJBPU07} Kawai, S., Bandrauk, A. D., Jaff\'{e}, C., Bartsch, T., Palacian, J., Uzer, T.: Transition state theory for laser-driven reactions. The Journal of chemical physics. {\bf 126(16)} 164306 (2007).

\bibitem{VP00} Vesel\'{y}, K., Podolsk\'{y}, J.:Chaos in a modified H\'{e}non-Heiles system describing geodesics in gravitational waves. Physics Letters A. {\bf 271(5-6)} 368-376 (2000).

\bibitem{VL96} Vieira, W. M., Letelier, P. S.: Chaos around a Hénon-Heiles-inspired exact perturbation of a black hole. Physical review letters. {\bf 76(9)} 1409 (1996).

\bibitem{V79} Verhulst, F.: Discrete symmetric dynamical systems at the main resonances with applications to axisymmetric galaxies. Philos. Trans. R. Soc. A. \textbf{290} 435-465 (1979).

\bibitem{ZRD18} Zotos, E. E., Riaño-Doncel, A., Dubeibe, F. L.: Basins of convergence of equilibrium points in the generalized Hénon–Heiles system. International Journal of Non-Linear Mechanics. {\bf 99} 218-228 (2018).

\bibitem{DRZ18} Dubeibe, F. L., Ria\~no-Doncel A., Zotos, E. E.: Dynamical analysis of bounded and unbounded orbits in a generalized H\'enon-Heiles system. Phys. Lett. A. \textbf{382} 904-910 (2018).

\bibitem{C02} Contopoulos, G.: Order and Chaos in Dynamical Astronomy, pp. 433-435. Springer, Berlin (2002).

\bibitem{L07} Lyapunov, A.M.: Probl\`{e}me g\'{e}n\'{e}ral de la stabilit\'{e} du movement. Ann. Fac. Sci. Toulouse, \textbf{9} 203-475 (1907).

\bibitem{L49} Lyapunov, A. M.: Annals of Mathematical Studies, Vol. 17, (1949).

\bibitem{S01} Skokos, C.: Alignment indices: a new, simple method for determining the ordered or chaotic nature of orbits. Journal of Physics A \textbf{34} 10029 (2001).

\bibitem{PTFV92} Press, W. H., Teukolsky, S. A., Flannery, B. P., Vetterling, W. T.: Numerical recipes in Fortran 77. Cambridge university press, Cambridge (1992).

\bibitem{KR92} Kaluza, M., Robnik, M.: Improved accuracy of the Birkhoff-Gustavson normal form and its convergence properties. J. Phys. A. \textbf{25} 5311 (1992).

\bibitem{B84} Barbanis, B.: The stochastic behaviour of a galactic model dynamical system. Celestial mechanics, \textbf{33(4)} 385-395 (1984).

\bibitem{DWGGS16} Daza, A., Wagemakers, A., Georgeot, B., Gu\'ery-Odelin, D., Sanju\'an, M. A.: Basin entropy: a new tool to analyze uncertainty in dynamical systems. Scientific reports, {\bf 6(1)} 1-10 (2016).

\end{thebibliography}
\end{document}